\documentclass[preprint,aps,prd,showpacs,nofootinbib,tightenlines]{revtex4-1}
\usepackage{amsmath}
\usepackage{bm}
\usepackage{times}
\usepackage{braket}
\usepackage{color,graphicx}
\usepackage{slashed}
\usepackage{hyperref}
\definecolor{Red}{rgb}{1.,0.,0.}

\definecolor{Blue}{rgb}{0.,0.,1.}

\definecolor{nicered}{rgb}{0.7,0.1,0.1}
\definecolor{nicegreen}{rgb}{0.1,0.5,0.1}

\hypersetup{colorlinks,citecolor=nicegreen,linkcolor=nicered}

\begin{document}

\newcommand{\beq}{\begin{eqnarray}}
\newcommand{\eeq}{\end{eqnarray}}
\newcommand{\non}{\nonumber\\ }

\newcommand{\jpsi}{J/\Psi}

\newcommand{\pka}{\phi_{K}^A}
\newcommand{\pkp}{\phi_K^P}
\newcommand{\pkt}{\phi_K^T}

\newcommand{\pea}{\phi_{\eta}^A}
\newcommand{\pep}{\phi_{\eta}^P}
\newcommand{\pet}{\phi_{\eta}^T}
\newcommand{\peqa}{\phi_{\eta_q}^A}
\newcommand{\peqp}{\phi_{\eta_q}^P}
\newcommand{\peqt}{\phi_{\eta_q}^T}

\newcommand{\pesa}{\phi_{\eta_s}^A}
\newcommand{\pesp}{\phi_{\eta_s}^P}
\newcommand{\pest}{\phi_{\eta_s}^T}

\newcommand{\pecv}{\phi_{\eta_c}^v}
\newcommand{\pecs}{\phi_{\eta_c}^s}

\newcommand{\pepa}{\phi_{\eta'}^A}
\newcommand{\pepp}{\phi_{\eta'}^P}
\newcommand{\pept}{\phi_{\eta'}^T}

\newcommand{\pksa}{\phi_{K^*}}
\newcommand{\pksp}{\phi_{K^*}^s}
\newcommand{\pkst}{\phi_{K^*}^t}

\newcommand{\fb}{f_B }
\newcommand{\fk}{f_K }
\newcommand{\fe}{f_{\eta} }
\newcommand{\fep}{f_{\eta'} }
\newcommand{\rks}{r_{K^*} }
\newcommand{\rk}{r_K }
\newcommand{\re}{r_{\eta} }
\newcommand{\rep}{r_{\eta'} }
\newcommand{\mb}{m_B }
\newcommand{\mtp}{m_{t'}}
\newcommand{\mw}{m_W }
\newcommand{\im}{{\rm Im} }
\newcommand{\kks}{K^{(*)}}
\newcommand{\etar}{\eta^{\prime}}
\newcommand{\acp}{{\cal A}_{CP}}

\newcommand{\etap}{\eta^{(\prime)} }
\newcommand{\pb}{\phi_B}
\newcommand{\pks}{\phi_{K^*}}

\newcommand{\xeba}{\bar{x}_2}
\newcommand{\xsba}{\bar{x}_3}
\newcommand{\res}{r_{\eta_s}}
\newcommand{\red}{r_{\eta_q}}
\newcommand{\peas}{\phi^A_{\eta_s}}
\newcommand{\peps}{\phi^P_{\eta_s}}
\newcommand{\pets}{\phi^T_{\eta_s}}
\newcommand{\pead}{\phi^A_{\eta_q}}
\newcommand{\pepd}{\phi^P_{\eta_q}}
\newcommand{\petd}{\phi^T_{\eta_q}}

\newcommand{\tp}{ t^\prime }
\newcommand{\bp}{ b^\prime }
\newcommand{\pvsl}{ p \hspace{-2.0truemm}/_{K^*} }
\newcommand{\esl}{ \epsilon \hspace{-2.1truemm}/ }
\newcommand{\psl}{ P \hspace{-2.4truemm}/ }
\newcommand{\nsl}{ n \hspace{-2.2truemm}/ }
\newcommand{\vsl}{ v \hspace{-2.2truemm}/ }
\newcommand{\epsl}{\epsilon \hspace{-1.8truemm}/\,  }
\newcommand{\bfkk}{{\bf k} }
\newcommand{\calm}{ {\cal M} }
\newcommand{\calh}{ {\cal H} }



\def \ahep{ {\bf Adv.High Energy Phys.} }
\def \cpc{ {\bf Chin. Phys. C} }
\def \ctp{ {\bf Commun.Theor.Phys. } }
\def \epjc{{\bf Eur.Phys.J. C} }
\def \jpg{ {\bf J.Phys. G} }
\def \npb{ {\bf Nucl.Phys. B} }
\def \plb{ {\bf Phys.Lett. B} }
\def \pr{  {\bf Phys. Rep.} }
\def \prd{ {\bf Phys.Rev. D} }
\def \prl{ {\bf Phys.Rev.Lett.}  }
\def \ptp{ {\bf Prog. Theor. Phys. }  }
\def \rmp{ {\bf Rev.Mod.Phys. }  }
\def \zpc{ {\bf Z.Phys.C}  }
\def \jhep{ {\bf J. High Energy Phys.}  }

\title{$B \to K \etap$ decays in the SM with fourth generation fermions}
\author{Shan Cheng, Ying-Ying Fan and Zhen-Jun Xiao\footnote{Electronic address: xiaozhenjun@njnu.edu.cn}}
\affiliation{Department of Physics and Institute of Theoretical Physics,
Nanjing Normal University, Nanjing, Jiangsu 210023, P.R.China }
\date{\today}
\begin{abstract}
By employing the perturbative QCD (pQCD) factorization approach,
we calculate the new physics contributions to the four
$B \to K \etap$ decays in the Standard Model (SM) with a fourth generation of fermions (SM4),
induced by the loop diagrams involving $\tp$ quark.
Within the considered parameter space of the SM4 we find that
(a) the next-to-leading order (NLO) pQCD predictions for the branching ratios and CP-violating asymmetries in
both the SM and SM4 generally agree with the data within one standard deviation;
(b) for $Br(B \to K \eta)$,  the inclusion of the fourth generation contributions can improve
the agreement between the theoretical predictions and the data effectively;
(c) for $Br(B \to K \etar)$, however, the decrease due to $\tp$ loops is disfavored by the data;
and (d) the new physics corrections to the CP-violating asymmetries of the considered decays
are about $10\%$ only.
\end{abstract}

\vspace{1cm}


\maketitle

\newpage
\section{Introduction}

As a simple extension of the standard model(SM), the standard model with the fourth generation fermion (SM4)
was rather popular in 1980s\cite{plb192,plb193,plb196,prl58}.
But unfortunately, the direct searches at the LHC experiments \cite{cms-2012a,atlas-2012a,prd86-074014}
have not found any signs of the heavy fourth generation $\tp$ and $\bp$ quarks so far.
The phenomenological studies  of the electroweak precision observables (EWPOs)\cite{prl81,prl65,pr330-263} and
some B meson rare decays \cite{prd82-033009,prd85-014008,plb683-302,prd84-014019,prd84-094027}
also resulted in some constraints on the parameter space of the SM4.
The observation of the SM Higgs boson at a mass of $126$GeV as reported by the CMS and ATLAS Collaboration \cite{atlas-1,cms-2} leads to very strong limits on the SM4:
it was claimed  \cite{prl109,lenz} that the SM4 was ruled out at $5.3\sigma$ by the Higgs data. 
But the authors of Ref.~\cite{he2012} also point out recently that
a SM4 with two-Higgs-doublets (4G-2HDM) can  explain current 126GeV Higgs signals.
The loop diagrams (box or penguins) involving the fourth generation fermions
$\tp$ and $\bp$, as is well-known, can provide new physics(NP) corrections to the branching ratios and CP violating asymmetries of B meson decays, such as the $B \to K \etap$ decays.
At present, it is still interesting to study the possible new physics effects to those well measured
B meson rare decays and to draw additional constraints on the SM4 from the relevant phenomenological analysis.
Such constraints are complimentary to those obtained from the EWPOs and/or the Higgs data.

The $B \to K \etap$ decays are penguin dominated decays, and have been studied intensively
by many authors for example in Refs.~\cite{kagan,xiao99,yy01,kou02,xnlo2008} in the framework of the SM
or various new physical models.
These four decays are studied very recently \cite{fan2013} by employing the perturbative QCD (pQCD)
factorization approach with the inclusion of all known next-to-leading order (NLO) contributions
from different sources, and the NLO pQCD predictions for both the branching ratios and
the CP violating asymmetries agree well with the precision experimental measurements \cite{hfag2012,pdg2012}.

In this paper, we will study the possible loop contributions induced by the heavy $\tp$ quark appeared in the SM4.
We will focus on the following points:
\begin{enumerate}
\item
Besides all the known NLO contributions already considered in Ref.~\cite{fan2013},
we here will consider the effects of the $\tp$ contributions to the relevant Wilson coefficients
as presented in Refs.~\cite{prd82-033009,plb683-302,prd84-014019} on the $B \to K \etap$
decays in the conventional Feldmann-Kroll-Stech  (FKS) $\eta-\etar$ mixing scheme \cite{fks98}.

\item
We will check the SM4 parameter-dependence of the pQCD predictions for the branching ratios and CP-violating
asymmetries, such as those $|\lambda_{t'}|, \phi_{t'}, m_{\tp}$ with the definition
of $V_{t'b}^{\ast}V_{t's}=|\lambda_{t'}|\exp[i\phi_{\tp}]$.

\end{enumerate}

The paper is organized as follows. In Sec.~II, we give a brief review for the pQCD factorization approach
and the SM4 model. In Sec.~III, we will make numerical calculations and present the numerical results.
A short summary will be given in the final section.


\section{ Theoretical framework}\label{sec:f-work}

For the charmless $B \to K \etap$ decays, the corresponding weak
effective Hamiltonian can be written as \cite{buras96}:
\beq
{\cal H}_{eff} &=& \frac{G_{F}}{\sqrt{2}}     \Bigg\{ V_{ub} V_{uq}^{\ast} \Big[
     C_{1}({\mu}) O^{u}_{1}({\mu})  +  C_{2}({\mu}) O^{u}_{2}({\mu})\Big]
  -V_{tb} V_{tq}^{\ast} \Big[{\sum\limits_{i=3}^{10}} C_{i}({\mu}) O_{i}({\mu})
  \Big ] \Bigg\} + \mbox{H.c.},
 \label{eq:heff}
\eeq
where $q=d,s$, $G_{F}=1.166 39\times 10^{-5} GeV^{-2}$ is the Fermi constant,
$O_{i}$ ($i=1,...,10$) are the local four-quark operators\cite{buras96}.
The Wilson coefficients $C_i$ in Eq.~(\ref{eq:heff}) and the corresponding renormalization group
evolution matrix are known currently at LO and NLO level \cite{buras96}.

In the B-rest frame, we assume that the light final state meson $M_2$ and $M_3$ ( here $M_i$ refers to $K$ or
$\etap$) is moving along the direction of $n=(1,0, {\bf{0}}_T)$ and $v=(0,1,{\bf{0}}_T)$, respectively.
Using the light-cone coordinates, the $B$ meson momentum $P_B$ and the two final state
mesons' momenta $P_2$ and $P_3$ (for $M_2$ and $M_3$ respectively) can be written as
\beq
P_B = \frac{M_B}{\sqrt{2}} (1,1,{\bf 0}_{\rm T}), \quad P_2 = \frac{M_B}{\sqrt{2}}(1-r_3^2,r^2_2,{\bf 0}_{\rm T}), \quad
P_3 = \frac{M_B}{\sqrt{2}} (r_3^2,1-r^2_2,{\bf 0}_{\rm T}),
\eeq
while the anti-quark momenta are chosen as
\beq
k_1 &=& \frac{m_B}{\sqrt{2}}\left (x_1,0,{\bf k}_{\rm 1T}\right ), \quad
k_2 =\frac{m_B}{\sqrt{2}}\left (x_2(1-r_3^2),x_2 r^2_2,{\bf k}_{\rm 2T} \right ),\non
k_3&=& \frac{m_B}{\sqrt{2}}\left (x_3 r_3^2,x_3(1-r_2^2),{\bf k}_{\rm 3T} \right).
\eeq
where $r_i=m_i/M_B$ with $m_i$ is the mass of meson $M_i$,  and $x_i$ refers to
the momentum fraction of the anti-quark in each meson.
After making the same integrations over the small components $k_1^-$, $k_2^-$, and $k_3^+$ as in Ref.~\cite{fan2013}
we obtain the decay amplitude conceptually
\beq
{\cal A}(B \to M_2M_3 ) &\sim
&\int\!\! d x_1 d x_2 d x_3 b_1 d b_1 b_2 d b_2 b_3 d b_3 \non &&
\cdot \mathrm{Tr} \left [ C(t) \Phi_B(x_1,b_1) \Phi_{M_2}(x_2,b_2)
\Phi_{M_3}(x_3, b_3) H(x_i, b_i, t) S_t(x_i)\, e^{-S(t)} \right ],
\quad \label{eq:a2}
\eeq
where $b_i$ is the conjugate space coordinate of $k_{iT}$. In above equation,
$C(t)$ is the Wilson coefficient evaluated at scale $t$,
$H(x_i, b_i, t)$ is the hard kernel, and $\Phi_B(x_1,b_1)$ and $\Phi_{M_i}(x_i,b_i)$ are the wave function.
The function $S_t(x_i)$ and $e^{-S(t)}$ are the threshold and $K_{T}$ Sudakov factors which suppresses
the soft dynamics effectively \cite{li2003}.

In pQCD approach, the $B$ meson is treated as a very good heavy-light system.
Following Ref.~\cite{lu2003}, we can write the wave function of B meson as the form of
\beq
\Phi_B= \frac{i}{\sqrt{2N_c}} (\psl_B +m_B) \gamma_5 \phi_B ({\bf k_1}).
\label{bmeson}
\eeq
Here we adopted the widely used B-meson distribution amplitude
\beq
\phi_B(x,b)&=& N_B x^2(1-x)^2\mathrm{exp} \left  [ -\frac{M_B^2\ x^2}{2 \omega_{b}^2} -\frac{1}{2} (\omega_{b} b)^2\right],
\label{phib}
\eeq
where the normalization factor $N_B$ depends on the value of $\omega_b$ and $f_B$ and defined through
the normalization relation $\int_0^1dx \; \phi_B(x,b=0)=f_B/(2\sqrt{6})$.
We here also take the shape parameter $\omega_{b}=0.4\pm 0.04$ GeV.
For the final state kaon and $\etap$ mesons, we use the same wave functions and distribution functions as
those used in Ref.~\cite{fan2013}.

In the SM4 model, the classic $3\times 3$ Cabibbo-Kobayashi-Maskawa (CKM) matrix is extended into
a $4 \times 4$ CKM-like mixing matrix \cite{buras2010}
\beq
U_{SM4}=\left( \begin{array}{cccc}
V_{ud} & V_{us} & V_{ub} & V_{ub'}\\
V_{cd} & V_{cs} & V_{cb} & V_{cb'}\\
V_{td} & V_{ts} & V_{tb} & V_{tb'}\\
V_{t'd} & V_{t's} & V_{t'b} & V_{t'b'}
   \end{array} \right),
\label{eq:V4}
\eeq
with the $t'$ and $b'$ denote the fourth generation up- and down-type quark.

In the SM4, the $\tp$ quark play a similar role as the top quark in the loop diagrams
and will provide  new physics terms, such as  $B_{0}(x_\tp)$, $C_{0}(x_\tp)$, $D_{0}(x_\tp)$ and $E_{0}(x_\tp)$,
to those relevant SM Inami-Lim functions $B_{0}(x_t)$, $C_{0}(x_t)$, $D_{0}(x_t)$ and $E_{0}(x_t)$ directly\cite{ptp65}.
When the new physics contributions are taken into account, the ordinary SM Wilson coefficients $C_i(M_W)$
will be changed accordingly.
In the SM4, one can generally write the Wilson coefficients as the combination of the SM part
and the additional fourth generation contribution\cite{buras2010}
\beq
C_i(m_W,m_{\tp})= C_i^{SM}(m_W) + C_i^{4G}(m_W,m_{\tp}).
\eeq
As mentioned in previous sections, the three new physics input parameters in the SM4 include
$\lambda_{t'}$, $\phi_{t'}$ and $\mtp$.

For the mixing scheme of $\eta-\eta^\prime$, we here use conventional FKS scheme \cite{fks98}
in the quark-flavor basis: $\eta_q= (u\bar u +d\bar d)/\sqrt{2}$ and $\eta_s=s\bar{s}$;
\beq
\left(\begin{array}{c} \eta \\ \eta^{\prime}
\end{array} \right) = \left(\begin{array}{c}
 F_1(\phi) (u\bar{u} + d\bar{d})  + F_2(\phi) \; s\bar{s} \\
 F_1^\prime(\phi) (u\bar{u} + d\bar{d})  + F_2^\prime(\phi) \; s\bar{s} \\
\end{array} \right)\label{eq:e-ep2}
\eeq
where $\phi$ is the mixing angle, and the mixing parameters are defined as
\beq
\sqrt{2}F_1(\phi)=F_2^\prime(\phi)=\cos(\phi), \non
F_2(\phi)=-\sqrt{2}F_1^\prime(\phi)=-\sin(\phi),
\label{eq:fiphi1}
\eeq
The relation between the decay constants $(f_\eta^q, f_\eta^s,f_{\etar}^q,f_{\etar}^s)$ and
$(f_q,f_s,)$ can be found in Ref.~\cite{xnlo2008}. The chiral enhancement $m_0^q$ and $m_0^s$ have been defined
in Ref.~\cite{nlo2005} by assuming the exact isospin symmetry $m_q=m_u=m_d$.
The three input parameters $f_q, f_s$ and $\phi$ in the FKS mixing scheme have been
extracted from the data of the relevant exclusive processes \cite{fks98}:
\beq
f_q=(1.07\pm 0.02)f_{\pi},\quad f_s=(1.34\pm 0.06)f_{\pi},\quad \phi=39.3^\circ\pm 1.0^\circ,
\eeq
with $f_\pi=0.13$ GeV.

\section{$B \to K \etap$ decays, the numerical results}\label{sec:lo-1}

\subsection{NLO contributions in the pQCD approach}

In Ref.~\cite{fan2013}, the authors studied the four $B \to K \etap$ decays with the inclusion
of all known NLO contributions by using the pQCD factorization approach.
For the SM part of the relevant decay amplitudes we use the formulaes as presented in
Ref.~\cite{fan2013}, where the authors confirmed numerically that
the still unknown NLO contributions from the relevant spectator and annihilation diagrams
are indeed small in size and can be neglected safely.

In this paper, we will take all known NLO contributions as considered in Ref.~\cite{fan2013} into account.
For the sake of the reader, we list these NLO contributions as follows:
\begin{enumerate}
\item[(1)]
The NLO Wilson coefficients $C_i(\mw)$ and the NLO renormalization group evolution matrix $U(t,m,\alpha)$
as defined in Ref.~\cite{buras96}, and the strong coupling constant $\alpha_s(t)$ at two-loop level.

\item[(2)]
The Feynman diagrams contributing to the hard kernel $H^{(1)}(\alpha_s^2)$ at the NLO level in the pQCD
approach include: (a) the Vertex Correction (VC) \cite{nlo2005}; (b) the Quark-Loop (QL) contributions\cite{npb675,nlo2005};
(c) the magnetic penguins (MG) contributions\cite{o8g2003,nlo2005};
and (d) the NLO part of the form factors (FF) as given in Ref.~\cite{prd85-074004}.

\end{enumerate}
For the explicit expressions of the decay amplitudes for the four $B \to K \etap$ decays and the
relevant functions, one can see Ref.~\cite{fan2013}. We here focus on the NP contributions from the heavy
$\tp$ quark.

\subsection{$Br(B \to K \etap)$ in SM4}\label{subs:br}

We use the following input parameters \cite{hfag2012,pdg2012} in the
numerical calculations(all masses and decay constants in units of GeV)
\beq
f_B &=& 0.21\pm 0.02, \quad f_K = 0.16, \quad m_{\eta}=0.548, \quad m_{\eta^{\prime}}=0.958,\non
m_{K^0} &=& 0.498, \quad m_{K^+} = 0.494, \quad m_{0K} = 1.7, \quad M_B = 5.28,\non
m_b &=& 4.8, \quad m_c = 1.5, M_W = 80.41, \quad \tau_{B^0} = 1.53 {\rm ps}, \quad \tau_{B^+} = 1.638 {\rm ps}.
\label{eq:para}
\eeq
For the CKM quark-mixing matrix in the SM, we adopt the Wolfenstein
parametrization as given in Ref.~\cite{hfag2012,pdg2012} and take
$ A = 0.832$, $\lambda = 0.2246$, $\bar{\rho} = 0.130 \pm 0.018$,
$\bar{\eta} = 0.350 \pm 0.013$.
For the three NP parameters, we choose similar values as in Ref.~\cite{plb683-302}:
\beq
|\lambda_{t'}|= 0.015\pm 0.010, \quad \phi_{t'}=0^0\pm 45^0, \quad \mtp = (600\pm 400) GeV.
\label{eq:newinputs}
\eeq

We here firstly calculate the branching ratios of the considered decay modes in both the SM and SM4
by employing the pQCD factorization approach.
In the B-rest frame, the branching ratio of a general $B \to M_2 M_3$ decay can be written as
\beq
Br(B\to M_2 M_3) = \tau_B\; \frac{1}{16\pi m_B}\; \chi\; \left | \calm(B\to M_2 M_3) \right|^2,
\label{eq:br-pp}
\eeq
where $\tau_B$ is the lifetime of the B meson,
$\chi\approx 1$ is the phase space factor and equals to unit when
the masses of final state light mesons are neglected.

When all currently known NLO contributions are included, we find the pQCD
predictions for $Br(B \to K \etap)$ in the SM4 (in unit of $10^{-6}$):
\beq
Br(B^0\to K^0 \eta)  &=& 1.46^{+0.30}_{-0.17}(\omega_b)^{+1.33}_{-0.57}(m_s)^{+0.28}_{-0.22}(f_B) ^{+0.53}_{-0.45}(a^{\eta}_2)
^{+0.03}_{-0.09}(|\lambda_{\tp}|)^{+0.14}_{-0.14}(\phi_{\tp})^{+0.04}_{-0.11}(\mtp),  \non
Br(B^0\to K^0 \etar) &=& 44.1^{+15.8}_{-10.3}(\omega_b)^{+11.6}_{-9.7}(m_s)^{+8.3}_{-8.3}(f_B)^{+1.2}_{-0.6}(a^{\eta}_2)
^{+3.1}_{-1.1}(|\lambda_{\tp}|)^{+3.5}_{-3.5}(\phi_{\tp})^{+4.0}_{-2.1}(\mtp),  \non
Br(B^+\to K^+ \eta) &=& 3.59^{+1.32}_{-1.02}(\omega_b)^{+2.37}_{-1.57}(m_s)^{+0.67}_{-0.68}(f_B)^{+0.88}_{-0.77}(a^{\eta}_2)
^{+0.23}_{-0.08}(|\lambda_{\tp}|)^{+0.34}_{-0.33}(\phi_{\tp})^{+0.38}_{-0.10}(\mtp),  \non
Br(B^+\to K^+ \etar) &=& 51.7^{+13.0}_{-9.8}(\omega_b)^{+12.6}_{-6.8}(m_s)^{+9.9}_{-9.7}(f_B)^{+2.2}_{-1.3}(a^{\eta}_2)
^{+1.5}_{-4.1}(|\lambda_{\tp}|)^{+4.3}_{-4.7}(\phi_{\tp})^{+1.9}_{-5.1}(\mtp),
\label{eq:br-np1}
\eeq
where the major theoretical errors are induced by the uncertainties of two sets of input parameters:
\begin{enumerate}
\item
The ordinary ``SM" input parameters: $\omega_b=0.4 \pm 0.04$ GeV, $m_s=0.13\pm 0.03$ GeV, $f_B=0.21\pm 0.02$ GeV
and Gegenbauer moment $a^{\eta}_2=0.44\pm 0.22$ ( here $a_2^\eta$ denotes $a_2^{\eta_q}$ or $a_2^{\eta_s}$ ) respectively;

\item
The new physics input parameters with the uncertainties as defined in Eq.~(13).

\end{enumerate}

\begin{table}[thb]
\begin{center}
\caption{ The NLO pQCD predictions for the branching ratios (in unit of
$10^{-6}$) in the framework of the SM (column two) and SM4 ( column three).
As a comparison, the QCDF predictions \cite{npb675} and the measured values \cite{hfag2012}
are also listed in the last two columns. }
\label{br1}
\vspace{0.2cm}
\begin{tabular}{l| c | c | c | c}
\hline\hline
Channel~~~ & NLO$^{\rm SM}$ &  NLO$^{\rm SM4}$ & QCDF\cite{npb675} & Data\cite{hfag2012} \\
\hline
$Br(B^0 \to K^0 \eta)$   &$2.53^{+3.6}_{-1.7}$   & $1.46^{+1.49}_{-0.78}({\rm SM})^{+0.15}_{-0.20}({\rm NP})$ &$1.1 ^{+2.4}_{-1.5}$   &$1.23^{+0.27}_{-0.24}$ \\
$Br(B^0 \to K^0 \etar)$  &$57.1^{+23.7}_{-17.0}$ & $44.1^{+21.3}_{-16.4}({\rm SM})^{+6.2}_{-4.2}({\rm NP})$ &$46.5^{+41.9}_{-22.0}$ &$66.1\pm 3.1$          \\
\hline
$Br(B^+ \to K^+\eta)$    &$3.94^{+3.8}_{-2.2}$   & $3.59^{+2.93}_{-2.14}({\rm SM})^{+0.56}_{-0.36}({\rm NP})$ &$1.9 ^{+3.0}_{-1.9}$   &$2.4^{+0.22}_{-0.21}$   \\
$Br(B^+ \to K^+\etar)$   &$58.6^{+24.0}_{-17.2}$ & $51.7^{+20.8}_{-15.4}({\rm SM})^{+4.9}_{-8.1}({\rm NP})$ &$49.1 ^{+45.2}_{-23.6}$&$71.1\pm2.6$            \\
\hline\hline
\end{tabular}
\end{center} \end{table}

In Table I, we list the NLO pQCD predictions in the framework
of the SM (column two) or  the SM4 (column three). In column four we show the
NLO SM predictions based on the QCD factorization (QCDF) approach as given in
Ref.~\cite{npb675}. And finally, the world averaged values of
experimental measurements \cite{hfag2012} are given in the last column.
The SM predictions in the column two of Table I agree perfectly with those as given in
Ref.~\cite{fan2013} when the ordinary FKS $\eta-\etar$ mixing scheme was employed.
The theoretical errors labeled with ``SM" or ``NP" denote the quadrature
combination of the theoretical errors from the uncertainties of two sets
of input parameters $(\omega_b,m_s,f_B,a^{\eta}_2)$
and $(|\lambda_{\tp}|, \phi_{\tp},\mtp)$, respectively.
From the numerical results as shown in Eq.~(\ref{eq:br-np1}) and Table I,
we find the following points:
\begin{enumerate}
\item
The pQCD predictions for $Br(B\to K \etap)$ become smaller than the SM ones after the inclusion
of the new physics contributions due to the destructive interference between the SM and NP
contributions, but they still agree with the measure values
within one standard deviation Since the theoretical errors are still large.

\item
For $Br(B^0 \to K^0 \eta)$ ( $Br(B^+ \to K^+ \eta)$), the NP decrease
of the central value of the pQCD prediction is about $40\%$ ($10\%$).
The agreement between the theoretical predictions for $Br(B \to K \eta)$
is improved effectively after the inclusion of NP contributions.

\item
For $Br(B^0 \to K^0 \etar)$ ( $Br(B^+ \to K^+ \etar)$), however, the NP decrease
is about $23\%$ ($12\%$), but such changes are disfavored by the data.

\end{enumerate}

Although the four $B\to K \etap$ decays are generally penguin-dominated decays,
the relative strength of the penguin part against the tree and/or other parts can be
rather different for different decay modes.
The explicit numerical calculations tell us that the penguin contribution
play a more important rule in $B^0 \to K^0 \eta$ decay than in other
three decay modes in consideration, the $\tp$-penguins consequently provide
a much larger modification to $Br(B^0 \to K^0 \eta)$ ( a decrease about $40\%$ )
than to other decays ( a decrease from $10\%$ to $23\%$ in magnitude).

\subsection{CP-violating asymmetries in SM4}

Now we turn to the CP-violating asymmetries of $B \to K \etap$ decays in pQCD approach.
For $B^\pm \to K^\pm \eta$ decays, there is a large direct CP asymmetry ($\acp^{dir}$ ),
due to the destructive interference between the penguin amplitude and the tree amplitude.
The NLO pQCD predictions for the direct CP asymmetries (in units of $10^{-2}$)
$\acp^{dir}(B^\pm \to K^\pm \eta)$ and $\acp^{dir}(B^\pm \to K^\pm \etar)$ in the SM (column two)
and the SM4 (column three)are listed in Table II, the QCDF predictions and the data as given in
Refs.~\cite{npb675,hfag2012} are also given in last two columns as a comparison.

\begin{table}[thb]
\begin{center}
\caption{ The pQCD predictions for the direct CP asymmetries (in units of $10^{-2}$)
of charged $B^\pm \to K^\pm \etap$ decays in the SM and SM4. }
\label{dcp1}
\vspace{0.2cm}
\begin{tabular}{ c |  c |  c | c | c} \hline \hline
 Mode&  NLO$^{\rm SM}$ &  NLO$^{\rm SM4}$ &QCDF\cite{npb675}&  Data\cite{hfag2012} \\ \hline
$\acp^{dir}(K^\pm \eta)$  &$-25.9^{+13.8}_{-17.4}$      &$-27.9^{+12.4}_{-10.5}({\rm SM})^{+8.6}_{-6.7}({\rm NP})$    &$-19^{+29}_{-30}$         &$-37\pm 8$       \\ \hline
$\acp^{dir}(K^\pm \etar)$ &$-4.3^{+2.0}_{-1.6}$      &$-4.6^{+2.0}_{-2.0}({\rm SM})^{+1.3}_{-0.4}({\rm NP})$   &$-9.0^{+10.6}_{-16.2}$&$1.3^{+1.6}_{-1.7}$  \\
\hline \hline
\end{tabular}\end{center}
\end{table}

As to the CP-violating asymmetries for the neutral decays $B^0 \to K^0 \etap$, the effects of $B^0-\bar{B}^0$ mixing should be
considered. The explicit formulae for the CP-violating asymmetries of
$B^0(\bar B^0) \to K^0 \etap$ decays can be found easily, for example, in
Ref.~\cite{fan2013}, we here make the numerical calculations and then show the NLO
pQCD predictions for the direct and mixing-induced CP asymmetries in Table III.
The theoretical errors labeled with ``SM" or ``NP" have specified previously.

\begin{table}[thb]
\begin{center}
\caption{ The pQCD predictions for the CP asymmetries (in units of $10^{-2}$) for
neutral $B^0 \to K^0 \etap$ decays in the SM and SM4, and the measured values
as given by HFAG \cite{hfag2012}.}
\label{cp2}
\vspace{0.2cm}
\begin{tabular}{l |  c | c | c} \hline \hline
Mode & NLO$^{\rm SM}$ & NLO$^{\rm SM4}$ & Data \cite{hfag2012}  \\\hline
$\acp^{dir}(B^0\to K^0_S \eta)$    &$-11.0^{+4.0}_{-3.9}$  &$-14.8^{+5.0}_{-5.1}({\rm SM})^{+1.3}_{-0.6}({\rm NP})$  &$-$  \\ \hline
$\acp^{mix}(B^0\to K^0_S \eta)$    &$ 65.9^{+3.3}_{-5.1}$
&$71.4^{+3.2}_{-1.6}({\rm SM})\pm 0.03({\rm NP}) $ &$-$  \\ \hline
$\acp^{dir}(B^0\to K^0_S \etar)$   &$3.5 \pm 0.3$
&$4.1^{+0.2}_{-0.3} ({\rm SM}) ^{+0.6}_{-0.3} ({\rm NP})$  &$1 \pm 9$  \\ \hline
$\acp^{mix}(B^0\to K^0_S \etar)$   &$69.8\pm 0.3$
&$70.5^{+0.1}_{-0.2}({\rm SM})\pm 0.2 ({\rm NP})$  &$64 \pm 11$  \\
\hline \hline
\end{tabular}
\end{center}
\end{table}

From the numerical results as listed in Table II and III, one can see that
the new physics effects on the pQCD predictions for the CP-violating asymmetries
of the considered four decays are generally much smaller than the theoretical errors.

\section{summary }

In this paper, we calculated the new physics contributions to the four $B \to K \etap$ decays in
the SM4. From our numerical calculations and phenomenological analysis, we find the following points
\begin{enumerate}
\item
In both the SM and SM4, the pQCD predictions for the branching ratios and CP-violating asymmetries
agree with the data within one standard deviation, of course, partially due to the still large
theoretical errors.

\item
For $Br(B^0 \to K^0 \eta)$ and $Br(B^+ \to K^+ \eta)$,  the NP decrease is about $40\%$ and $10\%$ respectively,
the agreement between the theoretical predictions and the data is improved effectively
after the inclusion of NP contributions.

\item
For $Br(B^0 \to K^0 \etar)$ and  $Br(B^+ \to K^+ \etar)$, however, the NP decrease
is about $23\%$ and $12\%$ respectively, but such changes are disfavored by the data.

\item
The new physics corrections  on the CP-violating asymmetries
of the considered decays are about $10\%$ only.
\end{enumerate}

\begin{acknowledgments}
This work is supported by the National Natural Science Foundation of China under the Grant
No.~11235005, and by the Project on Graduate Students¡¯ Education and Innovation of Jiangsu
Province, under Grant No. CXZZ13\_0391;

\end{acknowledgments}



\begin{thebibliography}{99}

\bibitem{plb192}
W.S. Hou, A. Soni, and H. Steger, \prl {\bf 59}, 1521 (1987).

\bibitem{plb193}
G. Eilam, J.L. Hewett, and T.G. Rizzo, \plb {\bf 193}, 533 (1987).

\bibitem{plb196}
W.S. Hou and A. Soni A, \plb {\bf 196}, 92 (1987).

\bibitem{prl58}
W.S. Hou, R.S. Wikkey, and A. Soni, \prl {\bf 58}, 1608 (1987).

\bibitem{cms-2012a} 	
S.~Chatrchyan {\it et al.}, CMS Collaboration, \prd {\bf 86}, 112003 (2012).

\bibitem{atlas-2012a}
S. Nektarijevic, for ATLAS Collaboration, PoS HQL2012, 2012-070.

\bibitem{prd86-074014} 	
O. Eberhardt, A. Lenz, A. Menzel, U. Nierste, and M. Wiebusch,
\prd {\bf 86}, 074014 (2012).

\bibitem{prl81}
Y. Fukuda, T. Hayakawa, E. Ichihara et al., \prl {\bf 81}, 1158 (1998).

\bibitem{prl65}
M.E. Peskin and T. Takeuchi, \prl {\bf 65}, 4279 (1990).

\bibitem{pr330-263}   
P.H. Frampton, P. Q. Hung and M. Sher,  \pr  {\bf 330}, 263 (2000);
H.J. He, N. Polonsky, and S. Su, \prd {\bf 64}, 053004 (2001).

\bibitem{prd82-033009}
A. Soni, A.K. Alok, A. Giri, R. Mohanta, and S. Nandi,
\prd {\bf 82}, 033009 (2010).

\bibitem{prd85-014008}
R. Mohanta and A. Giri,  \prd {\bf 85}, 014008 (2012).

\bibitem{plb683-302}
A.~Soni, A.K.~Alok, A.~Grri, R.~Mohanta, and S.~Nandi, \plb {\bf 683}, 302(2011).

\bibitem{prd84-014019}
R.~Mohanta, \prd {\bf 84}, 014019 (2011).

\bibitem{prd84-094027}
W.S.~Hou, M.~Kohda,and F.R.~Xu, \prd {\bf 84}, 094027 (2011).

\bibitem{atlas-1}
G. Aad et al., (ATLAS Collaboration), \plb, 2012,  {\bf 716}: 1

\bibitem{cms-2}
S. Chatrchyan et al., (CMS Collaboration), \plb, 2012,  {\bf 716}: 30.

\bibitem{prl109}
O. Eberhardt, G. Herbert, H. Lacker, A. Lenz, A. Menzel, U. Nierste and M. Wiebusch,
\prl {\bf 109}, 241802 (2012).

\bibitem{lenz}
A. Djouadi and A. Lenz, \plb {\bf 715}, 310 (2012);\\
A. Lenz, \ahep  Vol. 2013, 910275 (2013).

\bibitem{he2012}
N. Chen and H.J. He, \jhep  {\bf 1204}, 062(2012).

\bibitem{kagan}
A.L.~Kagan and A.A.~Petrov, hep-ph/9707354;\\
S.~Khalil and E. Kou, \prl {\bf 91}, 241602 (2003).

\bibitem{xiao99}
Z.J.~Xiao, C.S.~Li, and K.T.~Chao \prd {\bf 63}, 074005(2001);\\
Z.J.~Xiao, K.T.~Chao and C.S.~Li, \prd {\bf 65}, 114021(2002);\\
Z.J. Xiao and W.J. Zou, \prd {\bf 70}, 094008 (2004).

\bibitem{yy01}
M.Z.~Yang and Y.D.~Yang, \npb {\bf 609}, 469 (2001);\\
M.~Beneke and M.~Neubert, \npb {\bf 651}, 225 (2003).

\bibitem{kou02}
E.~Kou and A.~Sanda, \plb {\bf 525}, 240 (2002).

\bibitem{xnlo2008}
Z.J.~Xiao, Z.Q.~Zhang, X.~Liu, and L.B.~Guo, \prd {\bf 78}, 114001 (2008).

\bibitem{fan2013}
Y.Y.~Fan, W.F.~Wang, S.~Cheng, and Z.J.~Xiao, \prd {\bf 87}, 094003 (2013).

\bibitem{hfag2012}
Y.~Amhis {\it et al}., (Heavy Flavor Averaging Group), arXiv:1207.1158 [hep-ex] and online update at http://www.slac.stanford.edu/xorg/hfag.

\bibitem{pdg2012}
J.~Beringer {\it et al.} (Particle Data Group),  \prd {\bf 86}, 010001 (2012).



\bibitem{fks98}
Th.~Feldmann, P.~Kroll, and B.~Stech, \prd {\bf 58}, 114006 (1998); \plb {\bf 449}, 339 (1999).

\bibitem{buras96}
G.~Buchalla, A.J.~Buras, and M.E.~Lautenbacher, \rmp {\bf 68}, 1125 (1996).

\bibitem{li2003}
H.N.~Li,  Prog. Part. $\&$ Nucl. Phys. {\bf 51}, 85 (2003) and references therein.

\bibitem{lu2003}
C.D.~Lu, M.Z.~Yang, \epjc {\bf 28}, 515 (2003).

\bibitem{buras2010}
A.J.~Buras, B.~Duling, T.~Feldmann, \jhep {\bf 1009}, 106 (2010).

\bibitem{ptp65}
T.~Inami, and C.S.~Lim, \ptp {\bf 65} (1981) 297, {\bf 65} (1981) 1772.


\bibitem{nlo2005}
H.N.~Li, S.~Mishima, A.I.~Sanda, \prd {\bf 72}, 114005 (2005), and referrences therein.

\bibitem{npb675}
M.~Beneke and M.~Neubert, \npb {\bf 675}, 333 (2003).

\bibitem{o8g2003}
S.~Mishima and A.I.~Sanda, Prog. Theor. Phys. {\bf 110}, 549 (2003).

\bibitem{prd85-074004}
H.N.~Li, Y.L.~Shen, and Y.M.~Wang, \prd {\bf 85}, 074004 (2012).

\end{thebibliography}
\end{document}